\documentclass[aps,prl,notitlepage,reprint,superscriptaddress,nofootinbib]{revtex4-1}

\usepackage{graphicx}
\usepackage{dcolumn}
\usepackage{bm}
\usepackage{amsmath}
\usepackage{upgreek}
\usepackage[colorlinks,linkcolor=red,citecolor=blue,urlcolor=red]{hyperref}
\usepackage[utf8]{inputenc}
\usepackage[T1]{fontenc}
\usepackage{mathptmx}
\usepackage{ulem}
\usepackage[colorinlistoftodos]{todonotes}
\usepackage[mathlines]{lineno}
\renewcommand{\t}[1]{\mathrm{#1}}
\AtBeginDocument{%
	\newwrite\bibnotes
	\def\bibnotesext{Notes.bib}
	\immediate\openout\bibnotes=\jobname\bibnotesext
	\immediate\write\bibnotes{@CONTROL{REVTEX41Control}}
	\immediate\write\bibnotes{@CONTROL{%
			apsrev41Control,author="08",editor="1",pages="1",title="0",year="1"}}
	\if@filesw
	\immediate\write\@auxout{\string\citation{apsrev41Control}}%
	\fi
}%

\begin{document}
	
	\title{Quantum-limited imaging of a nanomechanical resonator with a spatial mode sorter}
	
	\author{M. E. Choi}
	\affiliation{Wyant College of Optical Sciences, University of Arizona, Tucson, AZ 85721, USA}
	
	\author{C. M. Pluchar}
	\affiliation{Wyant College of Optical Sciences, University of Arizona, Tucson, AZ 85721, USA}
	
	\author{W. He}
        \affiliation{Wyant College of Optical Sciences, University of Arizona, Tucson, AZ 85721, USA}

 	\author{S. Guha}
	\affiliation{\mbox{Department of Electrical and Computer Engineering, University of Maryland, College Park, MD 20742, USA}}
	
	\author{D. J. Wilson}
	\affiliation{Wyant College of Optical Sciences, University of Arizona, Tucson, AZ 85721, USA}
	
	\date{\today}
	\begin{abstract}
  		We explore the use of a spatial mode sorter to image a nanomechanical resonator, with the goal of studying the quantum limits of active imaging and extending the toolbox for optomechanical force sensing. In our experiment, we reflect a Gaussian laser beam from a vibrating nanoribbon and pass the reflected beam through a commercial spatial mode demultiplexer (Cailabs Proteus). The intensity in each demultiplexed channel depends on the mechanical mode shapes and encodes information about their displacement amplitudes. As a concrete demonstration, we monitor the angular displacement of the ribbon’s fundamental torsion mode by illuminating in the fundamental Hermite-Gauss mode (HG$_{00}$) and reading out in the HG$_{10}$ mode. We show that this technique permits readout of the ribbon’s torsional vibration with a precision near the quantum limit.  Our results highlight new opportunities at the interface of quantum imaging and quantum optomechanics.
	\end{abstract}
	
	\maketitle

Spatial mode demultiplexing (SPADE) has emerged as a powerful tool for extreme imaging applications ranging from exoplanet detection \cite{norris2020all} to superresolution microscopy \cite{tsang2017subdiffraction}. Motivations include the advance of low-cross-talk commercial technologies based on multi-plane light conversion (MPLC) \cite{fontaine2019laguerre} and the identification of protocols for traditional low-light imaging applications (e.g. ground-based telescopy), in which SPADE arises as an optimally efficient receiver~\cite{rouviere2023experimental, grace2022identifying}.

A canonical application of SPADE is tracking small distortions (such as displacement) of a laser beam~\cite{defienne2024advances}. As opposed to traditional cameras, which decode information in a pixel basis, SPADE receivers sort in a natural paraxial basis such as the Laguerre- or Hermite-Gauss (HG) modes \cite{delaubertQuantumMeasurementsSpatial2006a}.  The reduced modal support (typically $\sim$10) of this approach comes with the advantage of higher bandwidth and access to quantum-limited detection of the sorted light, including at the single photon level \cite{santamaria_single-photon_2024}. These trade-offs form the basis for recent quantum-enhanced \cite{grenapin2023superresolution,padilla2024quantum} or inspired \cite{lee2022quantum,tan2023quantum} SPADE protocols using classical \cite{lee2022quantum,tan2023quantum} and non-classical \cite{grenapin2023superresolution,padilla2024quantum} light.

Here we consider SPADE for dynamic imaging of a nanomechanical resonator, a measurement task rooted in practical applications~\cite{rugar1990atomic,sader2024data} and fundamental experiments~\cite{aspelmeyer2014cavity}, whose implementation has close parallels to laser beam tracking. To our knowledge, MPLC-based tracking of a nanomechanical resonator has not been reported in the literature; however, a recent study applied photonic-integrated SPADE to dynamic imaging of a levitated nanoparticle \cite{dinter2024towards}.  Both studies are timely, given the emergence of ultracoherent vacuum-levitated \cite{gonzalez2021levitodynamics} and tether nanomechanical resonators \cite{engelsen2024ultrahigh} for which the use of ``cavity-free" multi-mode imaging techniques is approaching quantum limits \cite{militaru2022,hao2024back}.  In these experiments, the resonator couples different spatial modes of a laser field, and the readout task becomes similar to spatial mode sorting.  

	\begin{figure}[t!]
		\includegraphics[width=1\columnwidth]{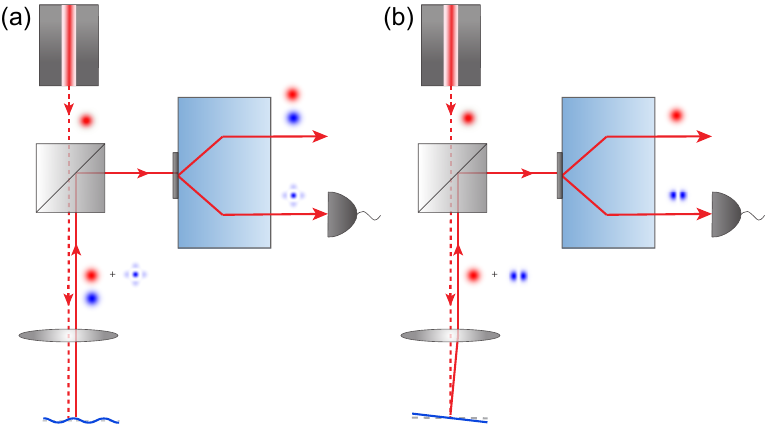}
		\caption{(a) Concept for readout of a mechanical oscillator with a spatial mode sorter. Different mode shapes represent optomechanical coupling between otherwise independent spatial modes.  Blue represents the fraction of light scattered to a different frequency. A mode sorter distills the orthogonal spatial mode. (b) Special case of an optical lever, in which angular displacement is encoded into a superposition of fundamental and first order Hermite-Gauss modes.  }
		\label{fig:1}
		\vspace{-2mm}
	\end{figure}

 \begin{figure*}[ht!]
		\vspace{-3mm}
		\includegraphics[width=2\columnwidth]{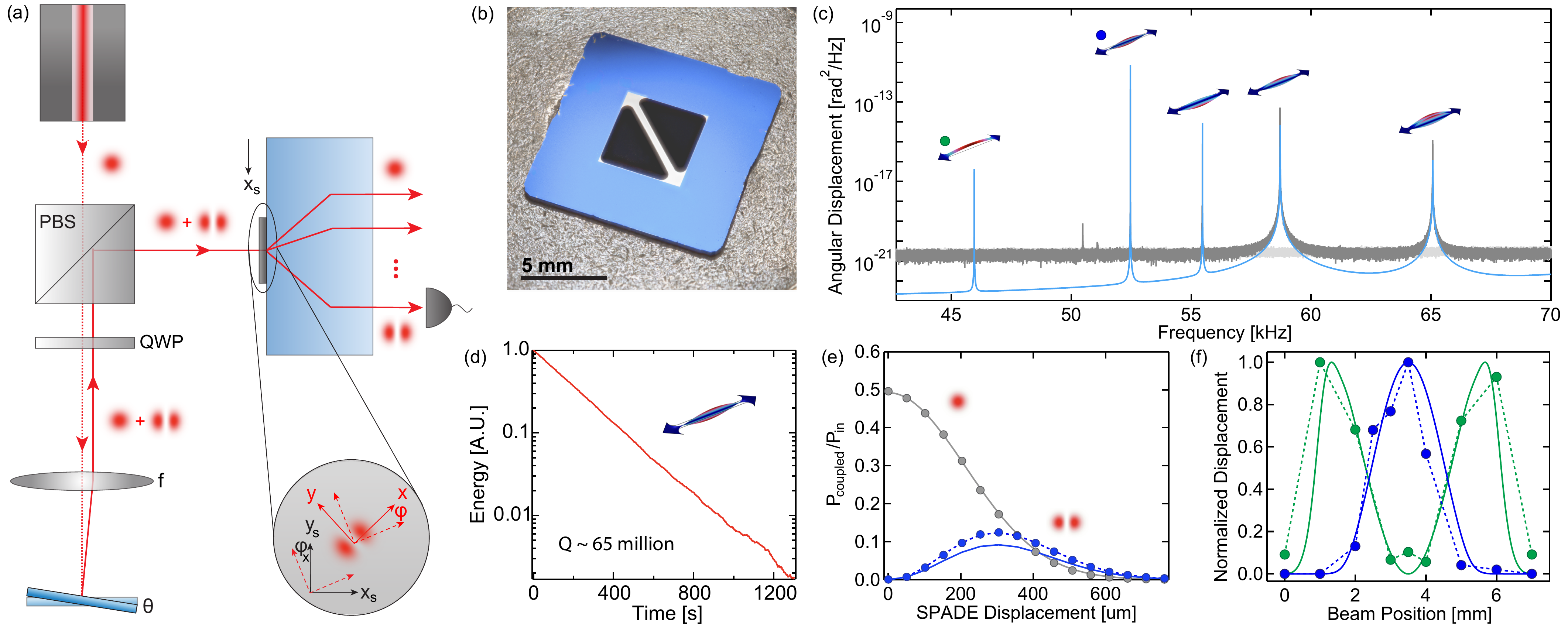}
		\caption{\textbf{SPADE-based optical lever readout of a nanomechanical torsion oscillator} (a) Sketch of the experiment: a nanoribbon is illuminated by a HG$_{00}$ laser beam.  Angular displacement $\theta$ of the ribbon is encoded in HG$_{10}$ component of reflected field, which is sorted using a HG-mode multi-plane light converter (MPLC). The coordinate frame of the torsion oscillator (red) is rotated with respect to the coordinate frame of the MPLC (black). (b) Photograph of the Si$_3$N$_4$ nanoribbon, 400 $\mu$m wide and 75 nm thick.
  (c) Broadband power spectral density of the HG$_{01}$ photocurrent calibrated in angular displacement units.  Blue curve is a multimode thermal noise model. Simulated modeshapes are assigned to each noise peak.
  (d) Ringdown of the fundamental torsion mode of the nanoribbon, revealing a quality factor $Q$=65 million.  (e) Power coupling efficiency of the reflected field into HG$_{00}$ and HG$_{10}$ ports of the MPLC, versus transverse misalignment $x$. (f) Normalized area of the fundamental flexural and torsion modes in panel (c) versus beam position along the ribbon axis (see main text for details).}  \label{fig:2}
		\vspace{-3mm}
	\end{figure*}

The basic experimental concept is illustrated in Fig. \ref{fig:1}. A laser beam is reflected from a vibrating membrane into a SPADE receiver.  The transverse modeshape of the reflected beam $u_\t{ref}(x,y)$ encodes the membrane displacement $z(x,y) = z_0\phi(x,y)$ into two orthogonal modes~\cite{pluchar2024imaging} 
\begin{equation}\label{eq:1}
u_\t{ref} = u_\t{in}e^{2ikz} \approx u_\t{in}+2ikz_0\left(\beta_\parallel u_\t{in}+\beta_\perp  u_\perp \right)
\end{equation}
where $u_\t{in}$ is the incident mode, $u_\perp$ is the orthogonal scattered mode, $\phi$ is the vibrational modeshape, $\beta_{\parallel(\perp)} = \langle \phi u_\t{in}|u_\t{in(\perp)}\rangle$ is a transverse overlap integral assuming $\langle u|u\rangle = 1$, and $k = 2\pi/\lambda$ is the laser wavenumber (for wavelength $\lambda$).  In the extreme case that $\beta_\parallel = 0$, displacement is fully encoded into the scattered mode and a straightforward calculation (see Appendix) predicts that SPADE-based direct detection of $u_\perp$ yields a quantum-limited displacement imprecision of 
\begin{equation}\label{eq:GeneralizedImprecision}
S_{z_0}^\t{imp} = \left(8 N k^2 \beta_\perp^2\right)^{-1} = \hbar^2/S_F^\t{BA}
\end{equation}
where $N$ is the mean reflected photon flux and $S_F^\t{BA}$ is the backaction force due to spatiotemporal photon shot noise \cite{pluchar2024imaging}.

To explore Eq. \ref{eq:GeneralizedImprecision}, we specialize to a $\t{HG}_{00}$ laser reflected from a torsion mode, $\phi\propto x$.  In this case, our setup embodies an optical lever (OL), which encodes angular displacement $\theta = z'(x)$ into a superposition of $\t{HG}_{00}$ and HG$_{10}$~\cite{hao2024back,enomoto2016SQLtorsion}
\begin{equation}\label{eq:2}
u_\t{ref} \approx  u_{00}+2i(\theta/\theta_\t{D}) u_{10}
\end{equation}
where $\theta_\t{D}$ is the diffraction angle of the incident beam.  When aligned into a $\t{HG}$ mode sorter and monitored through the HG$_{10}$ port, Eq. \ref{eq:GeneralizedImprecision} predicts a displacement imprecision of
\begin{equation}\label{eq:4}
S_\theta^\t{imp} = \frac{\theta_\t{D}^2}{8 N }
\end{equation}
which is the Heisenberg limit for an OL measurement subject to a quantum torque backaction of $S_\tau = 8\hbar^2 N \theta_\t{D}^2$ \cite{hao2024back,pluchar2024imaging,pluchar2024quantum}.

Our experimental setup is similar to that in~\cite{pluchar2024quantum}, where we studied the quantum limits of OL detection~\cite{pluchar2024quantum} using a Si$_3$N$_4$ nanoribbon with high-$Q$ torsion modes~\cite{pratt2023nanoscale} as a mechanical resonator and a split photodetector (SPD) as a receiver.  
For this study, we use an identical nanoribbon with a fundamental torsional resonance frequency of $f = 52.5$ kHz and $Q = 65\times10^6$.  The SPD is replaced with a commercial 10-channel HG-mode MPLC (Cailabs Proteus-C), which has a nominal throughput and cross-talk of $\sim 1\;\t{dB}$ and $\sim 20\;\t{dB}$, respectively, at an operating wavelength of $\lambda = 1550$ nm. As illustrated in Fig. \ref{fig:2}a, the ribbon is housed in a high vacuum chamber ($\sim10^{-8}\,\t{mbar}$) beneath the OL setup. The optical breadboard, collimation lens, and MPLC are mounted on individual micropositioning stages for fine alignment.

Alignment is a special consideration in our SPADE-based OL scheme for three main reasons. First, the reflected field must be mode-matched to the MPLC, otherwise HG$_{10}$ photons are lost to different channels.  Second, for $\theta/\theta_\t{D}\ll 1$, only a small fraction of the incident field is scattered to HG$_{10}$, rendering its direct detection sensitive to crosstalk. Third, the principle axes of the MPLC must be aligned to scattered field's.  
The first challenge can be eased by deriving the incident field from the MPLC---however, for technical reasons we illuminate from an approximately mode-matched auxiliary optical fiber. The second challenge can be addressed by laterally translating the MPLC to couple some $\t{HG}_{00}$ light into the HG$_{10}$ port, coherently amplifying the HG$_{10}$ signal at the cost of a reduction in sensitivity.  The last can be addressed by rotating either the sample or the MPLC.  Accounting for non-idealities in this alignment scheme, we consider the following sensitivity model (see Appendix and Fig. \ref{fig:2} for details)
\begin{equation}\label{eq:5}
	 	S_\theta^\t{imp} = \frac{\theta_\t{D}^2}{8N}\frac{1}{\eta_\t{d}}\frac{e^{x_\t{s}^2/w^2}\sec^2\varphi}{\left(1-\frac{x_\t{s}^2}{2 w^2}\left(1+\frac{\cos(\varphi-2\varphi_x)}{\cos\varphi}\right)\right)^2}\equiv \frac{\theta_\t{D}^2}{8N}\frac{1}{\eta}
\end{equation}
where $x_\t{s}$ is the lateral displacement of the MPLC and beam axes, $w$ is the beam waist on the MPLC, $\varphi$ and $\varphi_x$ are the angular misalignment of the MPLC from the sample and displacement axis, respectively, and $\eta_\t{d}\leq1$ is the combined loss due MPLC mode-mismatch and  detector quantum efficiency.

Measurements characterizing the alignment of our MPLC are shown in Fig. \ref{fig:2}e, implying access to a quantum efficiency of $\eta \gtrsim 50\%$.  Specifically, we record the power coupling efficiency of the reflected field $\eta_\t{d}^{ij}$ into the HG$_{00}$ and HG$_{10}$ ports and compare to the idealized (waist-matched) model~\cite{joyce1984alignment}
\begin{equation}\label{eq:modematchingcoefficients}
\begin{split}
    \eta_\t{d}^{00} &\approx \eta^0_{00} e^{-x^2/w^2}\\
    \eta_\t{d}^{10} &\approx \eta^0_{10} (x^2/w^2) e^{-x^2/w^2}\cos^2\varphi_x
\end{split}
\end{equation}
where $\eta^0_{ij}$ is an overall channel loss parameter.  Results agree qualitatively well with Eq. \ref{eq:modematchingcoefficients} using $w = 300\;\mu\t{m}$, $\varphi_x = 45^\circ$, $\eta_{00}^{0}= 50\%$, and $\eta_{10}^{0} = 67\%$ (see Appendix for details).

\begin{figure*}[ht!]
		\vspace{-2mm}
		\includegraphics[width=1.6\columnwidth]{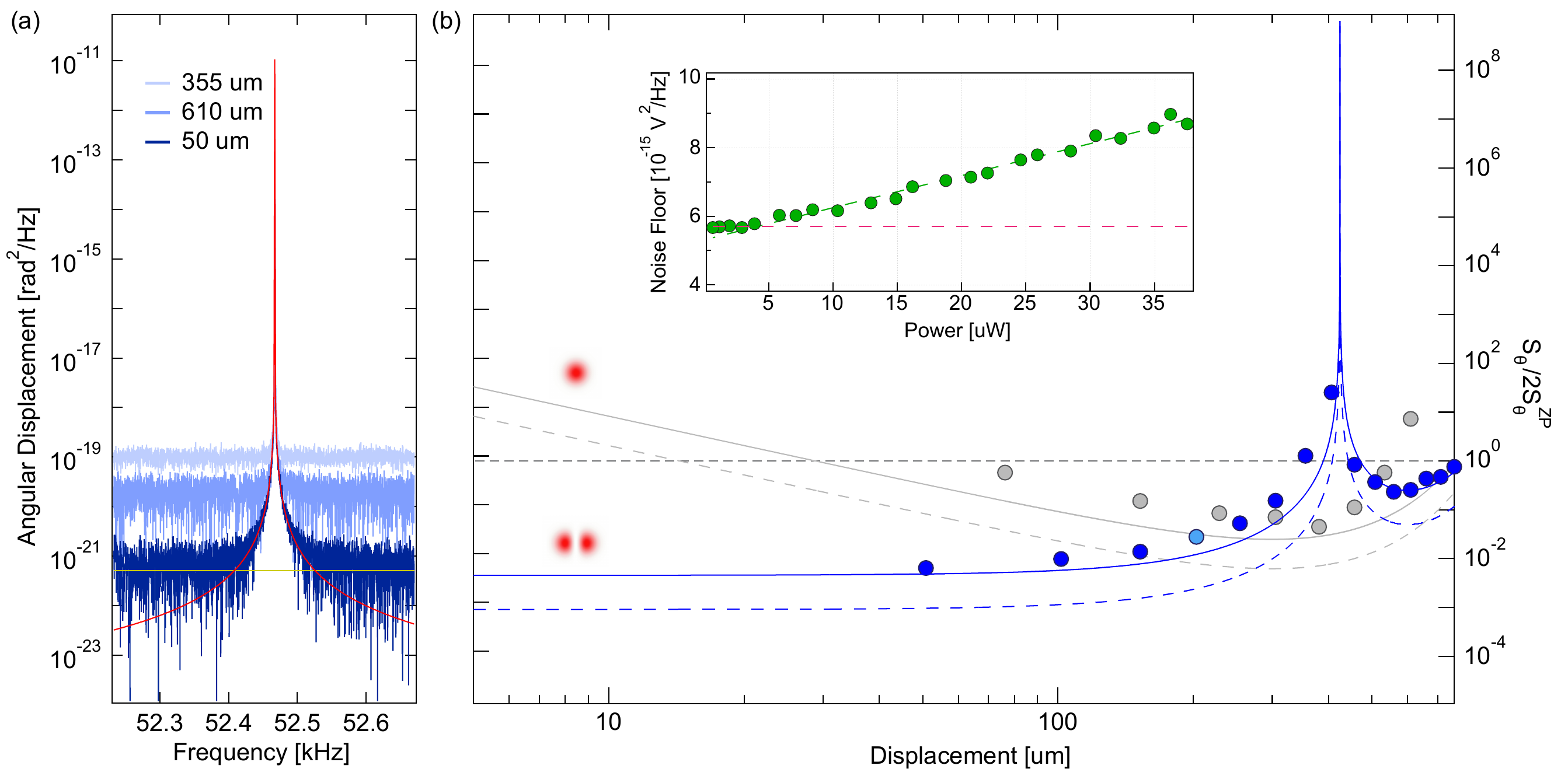}
		\caption{\textbf{Quantum-limited readout of a nanomechanical torsion oscillator with a spatial mode sorter.} (a) HG$_{10}$ photocurrent spectrum near the torsion mode resonance frequency for different MPLC lateral displacements $x_\t{s}$, calibrated to a thermal noise model (red). The inferred angular displacement imprecision is shown in yellow. 
 (a) Displacement imprecision versus $x_\t{s}$ (dots) compared to the model in Eq. \ref{eq:5} (lines).  Solid line is a fit to data. Dashed line assumes a perfect quantum efficiency $\eta = 1$.  Gray dots and curves are the corresponding measurements and model for the HG$_{00}$ photosignal (see Appendix). Inset: Raw HG$_{10}$ photocurrent noise versus power on the HG$_{10}$ photodetector.}\label{fig:3}
		\vspace{-2mm}
	\end{figure*}

Armed with these expectations, we carried out a series of experiments to explore the quantum efficiency of our SPADE-based OL, using the torsion mode of the nanoribbon in Fig. \ref{fig:2}b as a test mass.  For these experiments, the incident beam was focused to a spot size of $w_0\approx 150\;\mu\t{m}$ ($\theta_\t{D} = \lambda/(\pi w_0) = 3.3\;\t{mrad}$) on the ribbon, the reflected power was fixed at $P = 2.5\;\t{mW}$ ($N = P\lambda/hc \approx 2.0 \times 10^{16}$), and the MPLC was displaced by $x\gtrsim 100\;\mu\t{m}$ to amplify the signal into the HG$_{10}$ port above the photodetector noise.  The power spectrum of the HG$_{10}$ photocurrent was then recorded over a broad range of frequencies, using a low noise digitizer \cite{pratt2023nanoscale}. A representative measurement with $x \approx 200\;\mu\t{m}$ is shown in Fig. \ref{fig:2}c.  

We first remark that---as is typical in readout of high-$Q$ nanomechanical resonators---multiple thermal noise peaks appear in the broadband HG$_{10}$ photocurrent spectrum in Fig. \ref{fig:2}c, including the fundamental torsion mode at 52.5 kHz, with modeshape $\phi = (x/w_\t{r})\sin(\pi y/L)$, and fundamental flexural mode at 47 kHz, with modeshape $\phi = \sin(\pi y/L)$, where $w_\t{r} = 380\;\mu\t{m}$ and $L\approx 7\;\t{mm}$ are the ribbon width and length, respectively \cite{pluchar2024quantum}.  The flexural mode should in principle only couple to HG$_{10}$, implying a small MPLC orientational misalignment, $\varphi>0$.
To test this theory, we translated the beam position $(y_0)$ along the ribbon axis ($x = 0$) while monitoring the area $\langle \theta^2 \rangle$ beneath the torsion and flexural mode peaks, yielding qualitative agreement with the toy model $\langle \theta^2\rangle(y_0) \propto \beta_{01}^2(y_0)\cos^2\varphi + \beta_{10}^2(y_0)\sin^2\varphi \propto (\phi'_x(y_0))^2\cos^2\varphi +  (\phi'_x(y_0))^2 \sin^2\varphi$, valid for sufficiently small spot size, $w_0\ll w_\t{r},L_\t{r}$ (see Appendix). 

Without access to backaction (Eq. \ref{eq:2}), estimating the quantum efficiency of a displacement measurement requires direct calibration and comparison to a model.  To this end, following a standard approach, we calibrate
the HG$_{10}$ photocurrent spectrum in angular displacement units by bootstrapping the wings of the thermal noise peak at $\omega_\t{m} = 2\pi \times 52.5\;\t{kHz}$  to a narrowband model $S_\theta^\t{th}(\omega)\approx S_\theta^\t{th}/(1+4(\omega -\omega_\t{m})^2/\Gamma_\t{m}^2)$, where $S_\theta^\t{th} = 4k_B T_0 Q_\t{m}/I\omega_\t{m}$ is the resonant thermal displacement and $I = 2.8\times 10^{-18}\; \t{kg}\;\t{m}^2$ is the effective moment of inertia of the torsion mode (see Appendix).  We also carefully subtract the detector noise (measured by blocking the laser) and confirm that the remaining noise $S_\theta^\t{imp}$ is quantum noise by 
attenuating the laser and recording the raw voltage noise floor $S_V$, yielding characteristic shot noise scaling $S_V\propto P$ (Fig. \ref{fig:3}b inset).  

Fig. \ref{fig:3} summarizes our main result:  we recorded calibrated photocurrent spectra versus MPLC lateral displacement $x$, and observed imprecisions as low as $5\times 10^{-22}\;\t{rad}^2/\sqrt{\t{Hz}}$ at $x = 50\;\mu\t{m}$, corresponding to a quantum efficiency of $\eta \approx 14\%$ relative to the quantum-limited value of $\theta_\t{D}^2/(8N) = 7\times 10^{-23} \;\t{rad}^2/\sqrt{\t{Hz}}$ (Eq. \ref{eq:4}).  The solid (dashed) line in Fig. \ref{fig:3}b is an overlay of Eq. \ref{eq:5} with $\varphi_x = 45^\circ$, $\varphi = 0^\circ$, $w = 300\;\mu\t{m}$, and $\eta = 19\%$ ($100\%$, suggesting that a marginal increase in efficiency is possible with smaller $x$ (ultimately limited by cross-talk) and that $\sim 50\%$ of the loss is unaccounted for.

The observed high quantum efficiency and absolute sensitivity of our SPADE-based OL is independent of the small mass and high quality factor of the nanoribbon resonator.  Their combination signals access to spatio-temporal radiation pressure backaction (Eq. \ref{eq:GeneralizedImprecision}) and new opportunities at the interface of quantum imaging and quantum optomechanics \cite{pluchar2024imaging}.

To highlight this potential, a key figure of merit is the ratio of imprecision to the zero-point displacement spectral density of the oscillator, corresponding to the added measurement noise at the Standard Quantum Limit \cite{teufel2009nanomechanical}.  For the $x = 50\;\mu\t{m}$ measurement in Fig. \ref{fig:3}, the imprecision is 23 dB below the zero-point motion of the torsion mode $S_\theta^\t{ZP} = \hbar Q_\t{m}/(2I\omega_\t{m}^2) = 9\times 10^{-20}\;\t{rad}^2/\sqrt{\t{Hz}}$, corresponding to a phonon-equivalent imprecision and quantum backaction of $n_\t{imp} \equiv S_\theta^\t{imp}/2S_\theta^\t{ZP} = 0.003$ and $n_\t{BA} = (16 n_\t{imp}\eta)^{-1}\approx 150$, respectively \cite{wilson2015measurement}.

\begin{figure}[t!]
            \includegraphics[width=1\columnwidth]{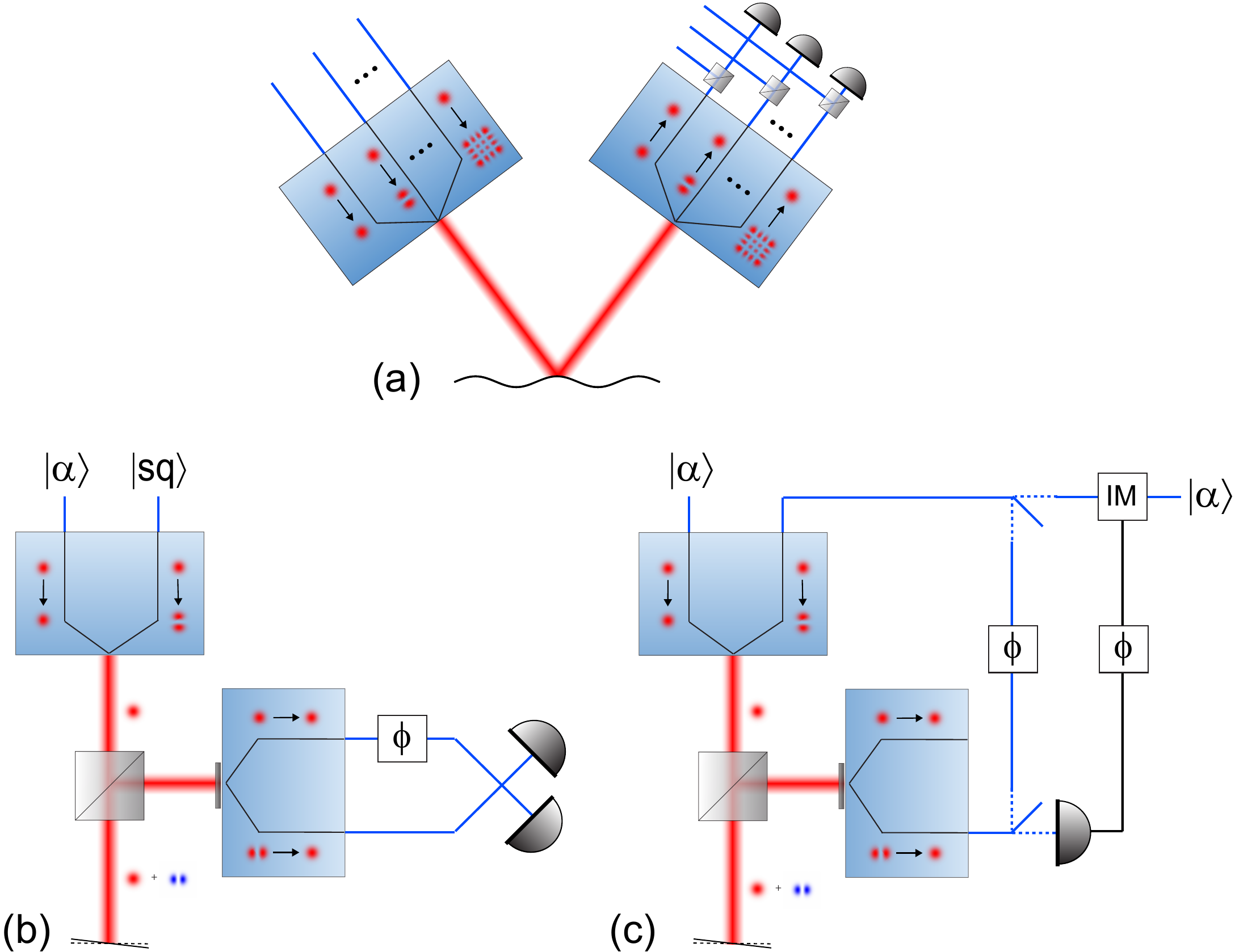}
		\caption{\textbf{Imaging-based quantum optomechanics using SPADE.} (a) Transceiver model for imaging of mechanical oscillator with a multimode transmitter and receiver produced by SPADE. (b,c) Examples of quantum-enhanced measurement and feedback control of a torsion oscillator with an HG-mode SPADE: $|\alpha\rangle$ (coherent state),  $|\t{sq}\rangle$ (squeezed state) $\phi$ (phase shifter), IM (intensity modulator).}
		\label{fig:4}
		\vspace{-2mm}
	\end{figure}

A canonical application of quantum-limited measurement in optomechanics is radiation pressure feedback cooling.  This might be implemented in our setup, for example, by imprinting the HG$_{10}$ photocurrent onto the HG$_{10}$ amplitude of the incident field using a pair of SPADEs, as shown in Fig. \ref{fig:4}c.  For ideal derivative feedback, a final occupation of $n_\t{m} = 2\sqrt{n_\t{imp}(n_\t{BA}+n_\t{th})}-0.5 \ge 0.5\left(1/\sqrt{\eta}-1\right)$~\cite{wilson2015measurement,rossi2018measurement} is achievable, where $n_\t{th} = k_B T_0/\hbar\omega_\t{m}$ is the bath phonon number.  For our nominal measurement efficiency and strength, $n_\t{m} = 1300$ is achievable from room temperature ($n_\t{th} = 1.2\times 10^8$) and $n_\t{m}= 0.9$ is achievable in the backaction limit ($n_\t{BA}\gg n_\t{th}$).   

Looking forward, we envision feedback cooling as one element in a broader program to improve the performance of SPADE-based displacement sensing, explore its radiation pressure quantum limits using nanomechanical resonators, and apply these findings to fundamental experiments and practical applications.  A transceiver model of this ``imaging-based" quantum optomechanics program is sketched in Fig. \ref{fig:4}a, in which a mechanical oscillator is imaged using a multimode transmitter and receiver based on a pair of SPADEs.  Using our base case of HG$_{00}$ illumination of a torsion oscillator as an example, Figs. \ref{fig:4}a,b show how quantum-enhanced readout, measurement-based feedback, or coherent feedback could be implemented, by injecting squeezed light, the measurement record, or the scattered field into the HG$_{10}$ (interacting mode) port of the input field, respectively.  These examples are readily extended to higher-order single- or multi-mode  squeezed light coupled to multi-mode mechanical resonators, providing a playground for testing entanglement-enhanced \cite{xia2023entanglement,treps2003quantumlaserpointer} and quantum-inspired \cite{lee2022quantum,he2024optimum} imaging protocols.  They also have a wide application space, from multi-mode ground-state preparation \cite{dinter2024towards} to label-free mass spectroscopy~\cite{sader2024data}.


\vspace{-2mm}
	\section*{Acknowledgements}
\vspace{-2mm}
 The authors thank Aman Agrawal for fabricating the devices; Andrew Land for engineering key elements of the optical setup; and Nico Deshler and Allison Rubenock, for helpful discussions. We also thank Mitul Dey Chowdhury for the photograph in Fig. \ref{fig:2}. This work was supported by the National Science Foundation (NSF) through award nos. 2239735 and 2330310. CMP acknowledges support from the ARCS Foundation. WH and SG acknowledge support from the Office of Naval Research (ONR) through contract no. N00014-19-1-2189, and the Air Force Office of Scientific Research 367 (AFOSR) through contract no. FA9550-22-1-0180. 

	\AtEndDocument{\include{SI_included}}
		\bibliography{ref}
	\end{document}


\title{Supplementary Information for\\``Quantum-limited imaging of a nanomechanical resonator with a spatial mode sorter"}
	
	\author{M. E. Choi}
	\affiliation{Wyant College of Optical Sciences, University of Arizona, Tucson, AZ 85721, USA}
	
	\author{C. M. Pluchar}
	\affiliation{Wyant College of Optical Sciences, University of Arizona, Tucson, AZ 85721, USA}
	
	\author{W. He}
        \affiliation{Wyant College of Optical Sciences, University of Arizona, Tucson, AZ 85721, USA}

 	\author{S. Guha}
	\affiliation{University of Maryland, College Park, MD 20742, USA}
	
	\author{D. J. Wilson}
	\affiliation{Wyant College of Optical Sciences, University of Arizona, Tucson, AZ 85721, USA}
	
	\date{\today}
	\begin{abstract}
  	Here we elaborate on various aspects of the theory, experiment, and data analysis described in the main text.
	\end{abstract}
	
	\maketitle
\vspace{-3mm}\section{Imprecision of SPADE-based displacement measurement}
To derive Eq. 2, we consider the setup in Fig. 1, where the photon flux in the scattered field $N_\perp$ is monitored with a lossless SPADE followed by a photodetector.   Using Eq. 1, 
\begin{equation}
    N_\perp = 4 N k^2 \beta_\perp^2 z_0^2. 
\end{equation} 

Combining the linear displacement sensitivity $\partial N_\perp/\partial z_0$ and the shot noise spectral density $S^\t{shot}_{N_\perp} = 2N_\perp$ yields the shot-noise-equivalent displacement (imprecision) given in Eq. 2:
\begin{equation}
    S^\t{imp}_{z_0} = \left( \frac{\partial N_\perp}{\partial z_0} \right)^{-2} S_{N_\perp}^\t{shot} =  \frac{ 1}{8 N k^2 \beta_\perp^2}.
\end{equation}

The above, linear response derivation of $S^\t{imp}_{z_0}$ assumes that $\partial N_\perp/\partial z_0$ is constant for small fluctuations in $z_0$ (on the order of picometers in our experiment).  In practice, this is achieved by slightly disaligning the SPADE receiver.

The right hand side of Eq. 2 is obtained from the expression for the generalized radiation pressure shot noise force \cite{pluchar2024imaging}
\begin{equation}
S^\t{BA}_{F} = 8\hbar^2Nk^2(\beta_\parallel^2+ \beta_\perp^2)\equiv8\hbar^2Nk^2\beta^2
\end{equation}
where 
\begin{equation}
\beta^2 = \iint u^2_\t{in}(x,y)\phi^2(x,y) dxdy  \equiv \langle u_\t{in}\phi | u_\t{in}\phi\rangle 
\end{equation}
is the transverse intensity overlap between the incident optical modeshape $u_\t{in}$ and the mechanical modeshape $\phi$, normalized such that $\langle u_\t{in}|u_\t{in}\rangle = 1$ and Max$(\phi)=1$.  Derivations of Eq. S3 are given in \cite{pluchar2024imaging,pinardEffectiveMassQuantum} and in the appendix of \cite{pluchar2024quantum}.

\vspace{-3mm}\section{HG-SPADE readout of a torsion oscillator}

In the main text, we specialize to HG$_{00}$ illumination of a torsion oscillator, which encodes angular displacement into the HG$_{10}$ mode of the reflected field.  Explicitly, we consider
\begin{subequations}\begin{align}
u_\t{in}(x,y)  = u_{00}(x,y) &= \sqrt{\frac{2}{\pi w_0^2}}e^{-(x^2+y^2)/w_0^2} \\
u_{10}(x,y) & = \frac{2x}{w_0} u_{00}(x,y)\\
\phi(x,y) & = \frac{2x}{w_\t{r}} \cos(\pi y/L)
\end{align}\end{subequations}
where $w_0$ is the beam waist and $w_\t{r}$ and $L$ are the width and length of our ribbonlike torsion oscillator, respectively.  In the limit that the ribbon is much larger than the incident beam $w,L\gg w_0$, the modeshape of the reflected field becomes
\begin{subequations}\begin{align}
u_\t{rel}(x,y) & = u_\t{in}(x,y)e^{2i k z_0\phi (x,y)}\\
&\approx u_{00}(x,y)\left(1+2ikz_0\phi(x,0)\right)\\
&=  u_{00}(x,y)+2i(k z_0 w_0/w_\t{r}) u_{10}(x,y)\\
&= u_{00}(x,y)+2i(\theta/\theta_\t{D}) u_{10}(x,y),
\end{align}\end{subequations}
where here we have identified $\theta_\t{D} = 2/(kw_0)$ as the diffraction angle of the incident beam and  $\theta = 2z_0/w_\t{r}$ as the angular displacement of the torsion mode.

As a check on Eq. S6c, it may be confirmed that
\begin{equation}
\beta_\perp = \langle u_{10}|\phi u_{00}\rangle  = w_0/w_\t{r}
\end{equation}
in the limit that $w,L\gg w_0$.  Equation S2 then yields Eq. 4:
\begin{equation}
    S_\theta^\t{imp} = \left(\frac{2}{w_\t{r}}\right)^2 S_{z_0}^\t{imp} = \frac{\theta_\t{D}^2}{8N}.
\end{equation}

\vspace{-3mm}\section{Inefficiency due to misalignment}

Misalignment of the SPADE receiver leads to measurement inefficiency, $\eta$.  We model this in Eq. 5 by considering the photon flux $N_\t{det}$ into the detection mode $u_\t{det}$
\begin{equation}
N_\t{det}(z_0) = N|\langle u_\t{det}|u_\t{ref}(z_0)\rangle|^2 
\end{equation}
and comparing Eq. S2 to 
\begin{equation}
    S^\t{imp,det}_{z_0} = \left( \frac{\partial N_\t{det}}{\partial z_0} \right)^{-2} S_{N_\t{det}}^\t{shot}.
\end{equation}
We thus find, assuming $k z_0\ll 1$, that
\begin{equation}
    \eta = \frac{S^\t{imp}_{z_0}}{S^\t{imp,det}_{z_0}}
    \approx |\langle u_\t{det}|u_\perp\rangle|^2.
\end{equation}

For lateral and angular misalignment as specified in Fig. 2, the HG$_{00}$ and HG$_{10}$ modes of the receiver can be modeled as
\begin{subequations}\begin{align}
u_{\t{det}}^{00} &= u_{00}(x-x_s\cos\phi_x,y-x_s\sin\phi_x) \\
u_{\t{det}}^{01} &=  u_{10}(x-x_s\cos\phi_x,y-x_s\sin\phi_x) \cos\phi\\&\;+  u_{01}(x-x_s\cos\phi_x,y-x_s\sin\phi_x)\sin\phi.
\end{align}\end{subequations}
Equation 5 is obtained by setting $u_\t{det}=u_{10}$ and assuming that the 
waist of the reflected and detection mode is $w$.

\vspace{-3mm}\section{HG$_{00}$ readout of the torsion mode}

Gray curves in Fig. 3b correspond to the displacement sensitivity when reading out in the HG$_{00}$ port.  We obtain these model curves by combining Eq. S10 and Eq. S12a, viz.
\begin{equation}
S_\theta^\t{imp,00} = \left(\frac{2}{w_\t{r}}\right)^2 \left(\frac{\partial N_{00}}{\partial z_0}\right)^{-2}S_{N_{00}^\t{shot}},
\end{equation}
with 
\begin{equation}
N_{00} = |\langle u_\t{det}^{00}|u_\t{ref}(z_0)\rangle|^2. 
\end{equation}

\vspace{-3mm}\section{Optomechanical coupling}

The model curves in Fig. 2f were computed by taking the derivative of finite-element-simulated (in COMSOL) modeshapes $\phi$ and noting that  for an incident beam with sufficiently small spot size translated along on the torsion axis ($x=0$), the optomechanical coupling at position $y=y_0$ along the torsion axis is approximated by
\begin{subequations}\begin{align}
    \beta_{10}(y_0)= \langle u_{10}(x,y)\phi(x,y-y_0)u_{00}(x,y)\rangle &\propto \frac{\partial\phi}{\partial x}\big|_{x=0, y=y_0}\\
   \beta_{01}(y_0)= \langle u_{01}(x,y)\phi(x,y-y_0)u_{00}(x,y)\rangle &\propto \frac{\partial\phi}{\partial y}\big|_{x=0, y=y_0}.
\end{align}\end{subequations}

\vspace{-3mm}\section{Spot size of the laser beam on the nanoribbon}
The spot size of the laser beam on the nanoribbon was estimated using a reflective knife-edge measurement as shown in Fig. S1, where the ribbon is the knife-edge.  The position of the collimating lens was adjusted to minimize the spot size and ensure that the ribbon coincided with the beam waist $w_0$.

\begin{figure}[h!]
\includegraphics[width=1\columnwidth]{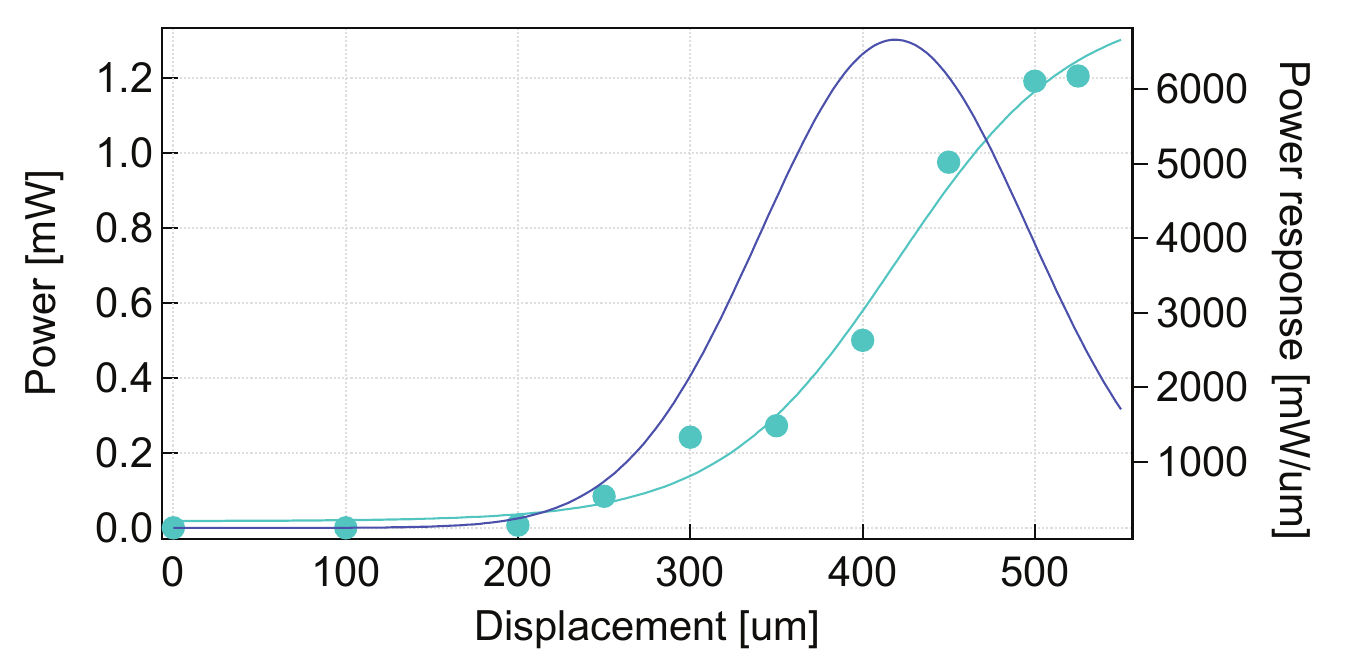}
	\caption{Reflective knife-edge measurement used to estimate the waist size $w_0$ of the incident beam and ensure it coincides with the nanoribbon.  From the measurement shown, we infer $w_0 \approx 150\;\mu\t{m}$.}
	\label{SI:1}
	\vspace{-3mm}
\end{figure}

\newpage

\bibliography{ref}